# Maximum Possibility Of Spectrum Access In Cognitive Radio Using Fuzzy Logic System

## K.Gowrishankar [1], Dr.C.Chandrasekar [2], R.Kaniezhil [3]

[1] Research Scholar, Department of Computer Science, Periyar University, Salem.

[2] Associate Professor, Department of Computer Science, Periyar University, Salem.

[3] Ph.D Research Scholar, Department of Computer Science, Periyar University, Salem.

**ABSTRACT**

Many spectrum methods have been proposed to use spectrum effectively, the opportunistic spectrum access has become the most viable approach to achieve near-optimal spectrum utilization by allowing secondary users to sense and access available spectrum opportunistically. However, a naive spectrum access for secondary users can make spectrum utilization inefficient and increase interference to adjacent users. In this paper, we propose a novel approach using Fuzzy Logic System (FLS) to control the spectrum access. We have to minimize the call blocking and interference. The linguistic knowledge of spectrum access are based on three descriptors namely spectrum utilization efficiency of the secondary user, its degree of mobility, and its distance to the primary user. The spectrum is chosen for accessing based on the maximum possibility for better utilization of the spectrum

## I. INTRODUCTION

In recent studies, the spectrum allocated by the traditional approach shows that the spectrum allocated to the primary user is under-utilized and the demand for accessing the limited spectrum is growing increasingly. Spectrum is no longer sufficiently available, because it has been assigned to primary users that own the privileges to their assigned spectrum. However, it is not used efficiently most of the time. In order to use the spectrum in an opportunistic manner and to the increase spectrum availability, the unlicensed users can be allowed to utilize licensed bands of licensed users, without causing any interference with the assigned service. The reason for allowing the unlicensed users to utilize licensed bands of licensed users if they would not cause any interference with the assigned service. This paradigm for wireless communication is known as opportunistic spectrum access and this is considered to be a feature of Cognitive Radio (CR).

Cognitive radio is an emerging wireless communication paradigm. The main challenge to opportunistic radio spectrum access lies in finding balance in conflicting goals of satisfying performance requirements for secondary user (SU) while minimizing interference to the active primary users (PU) and other secondary users. Secondary user should not degrade performance statistics of licensed primary users. In order to achieve these tasks, secondary user is required to recognize primary users, determine environment characteristics and quickly adapt its system parameters corresponding to the operating environment. Main abilities of cognitive radio (CR) with opportunistic radio spectrum access capabilities are spectrum sensing, dynamic frequency selection and adaptive transmit power control.

This paper presents a novel approach using Fuzzy logic system to utilize the available spectrum by the secondary users without interfering the primary user.

The paper is organized as follows; Section 2 and 3 defines cognitive radio and fuzzy logic system for its implementation. In section 4, opportunistic spectrum access by Fuzzy logic system to improve the spectrum efficiency. Section 5 and 6 presents the Knowledge Processing with Opportunistic Spectrum Access and simulation results. Finally, conclusions are presented in Section 7.

## II. COGNITIVE RADIO

The idea of cognitive radio was first presented officially in an article by Joseph Mitola III and Gerald Q. Maguire, Jr in 1999. Regulatory bodies in various countries found that most of the radio frequency spectrum was inefficiently utilized. For example, cellular network bands are overloaded in most parts of the world, but amateur radio and paging frequencies are not. This can be eradicated using the dynamic spectrum access.

The key enabling technology of dynamic spectrum access techniques is cognitive radio (CR) technology, which provides the capability to share the wireless channel with licensed users in an opportunistic manner. From this definition, two main characteristics of the cognitive radio can be defined as follows:

➢ Cognitive capability: It refers to the ability of the radio technology to capture or sense the information from its radio environment. Through this capability, the portions of the spectrum that are unused at a specific time or location can be identified. Consequently, the best spectrum and appropriate operating parameters can be selected.
- Reconfigurability: The cognitive capability provides spectrum awareness whereas reconfigurability enables the radio to be dynamically programmed according to the radio environment.

A CR "monitors its own performance continuously", in addition to "reading the radio's outputs"; it then uses this information to "determine the RF environment, channel conditions, link performance, etc.", and adjusts the "radio's settings to deliver the required quality of service subject to an appropriate combination of user requirements, operational limitations, and regulatory constraints".

Independent studies performed in some countries confirmed that observation and concluded that spectrum utilization depends strongly on time and place. Moreover, fixed spectrum allocation prevents rarely used frequencies (those assigned to specific services) from being used by unlicensed users, even when their transmissions would not interfere at all with the assigned service.

More specifically, the cognitive radio technology will enable the users to (1) determine which portions of the spectrum is available and detect the presence of licensed users when a user operates in a licensed band (spectrum sensing), (2) select the best available channel (spectrum management), (3) coordinate access to this channel with other users (spectrum sharing), and (4) vacate the channel when a licensed user is detected (spectrum mobility).

## III. FUZZY LOGIC SYSTEM

A fuzzy logic system (FLS) is unique in that it is able to simultaneously handle numerical data and linguistic knowledge. It is a nonlinear mapping of an input data (feature) vector into a scalar output, i.e., it maps numbers into numbers.

Fuzzy sets theory is an excellent mathematical tool to handle the uncertainty arising due to vagueness. Figure 1 shows the structure of a fuzzy logic system.

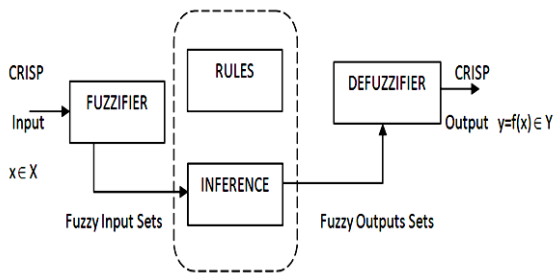

**Fig. 1.** The Structure of the Fuzzy Logic System

Since there is a need to "fuzzify" the fuzzy results we generate through a fuzzy system analysis i.e., we may eventually find a need to convert the fuzzy results to crisp results. Here, we may want to transform a fuzzy partition or pattern into a crisp partition or pattern; in control we may want to give a single-valued input instead of a fuzzy input command. The "dufuzzification" has the result of reducing a fuzzy set to a crisp single-valued quantity, or to a crisp set.

Consider a fuzzy logic system with a rule base of M rules, and let the lth rule be denoted by $R_l$. Let each rule have p antecedents and one consequent (as is well known, a rule with q consequents can be decomposed into rules, each having the same antecedents and one different consequent), i.e., it is of the general form

$R_l$ : IF $u_1$ is $F_l^1$ and $u_2$ is $F_l^2$ and … and $u_p$ is $F_l^p$, THEN $v$ is $G^l$.

where $u_k, K = 1,\ldots p$ and v are the input and output linguistic variables, respectively. Each $F_k^l$ and $G^l$ are subsets of possibly different universes of discourse. Let $F_k^l \subset U_k$ and $G^l \subset V$. Each rule can be viewed as a fuzzy relation $R_l$ from U to a set V where U is the Cartesian product $U = U_1,\ldots,U_p$. $R_l$ itself is a subset of the Cartesian product U X V = $\{(x, y): x \in U, y \in V\}$, where $x \equiv (x_1, x_2,\ldots,x_p)$ and $x_k$ and y are the points in the universes of discourse $U_k$ and V of $u_k$ and v.

While applying a singleton fuzzification, when an input $X' = \{x_1', x_2', x_3',\ldots x_p'\}$ is applied, the degree of firing corresponding to the lth rule is given by

$$x^* = \frac{\sum_{i=1}^{n} x_i \cdot \mu(x_i)}{\sum_{i=1}^{n} \mu(x_i)}$$

(1)

Where * denotes a T-norm, n represents the number of elements, $x_i$'s are the elements and $\mu(x_i)$ is its membership function. There are many kinds of dufuzzification methods, but we have chosen the centre of sets method for illustrative purpose. It computes a crisp output for the FLS by first computing the centroid, $C_{G^l}$ of every consequent set $G^l$ and, then computing weighted average of these centroids. The weight corresponding to the $l^{th}$ rule consequent centroid is the degree of firing associated with the $l^{th}$ rule, $T_{i=1}^{p} \mu_{F_l^i}(x_1')$, so that

$$y_{cos}(x') = \frac{\sum_{l=1}^{M} C_{G^l} T_{i=1}^{p} \mu_{F_1^l}(x_1')}{\sum_{l=1}^{M} T_{i=1}^{p} \mu_{F_1^l}(x_1')} \quad (2)$$

where M is the number of rules in the FLS.

## IV. OPPORTUNISTIC SPECTRUM ACCESS

We design the fuzzy logic for opportunistic spectrum access using cognitive radio. In this paper, we are selecting the best suitable secondary users to access the available users without any interference with the primary users. This is collected based on the following three antecedents i.e., descriptors. They are

Antecedent 1: Spectrum Utilization Efficiency
Antecedent 2: Degree of Mobility
Antecedent 3: Distance of Secondary user to the PU.

Fuzzy logic is used because it is a multi-valued logic and many input parameters can be considered to take the decision. Generally, the secondary user with the furthest distance to the primary user or the secondary user with maximum spectrum utilization efficiency can be chosen to access spectrum under the constraint that no interference is created for the primary user. In our approach, we combine the three antecedents to allocate spectrum opportunistically inorder to find out the optimal solutions using the fuzzy logic system.

Mobility of the secondary user plays a vital role in the proposed work. Wireless systems also differ in the amount of mobility that they have to allow for the users. Spectrum Mobility is defined as the process when a cognitive radio user exchanges its frequency of operation. The movement of the secondary user leads to a shift of the received frequency, called the Doppler shift.

Detecting signal from the primary user can be reduced using the Mobility. The secondary user should detect the primary signal which determines the spectrum that is unused. If the secondary user fails to detect the primary signal, then it will not determine exactly the spectrum that is unused. Thereby it leads to interference to the adjacent users. This is referred as the hidden node problem.

Besides, we consider the distance between the primary user and the secondary users. Distance between the primary user and the secondary users were calculated using the formula

$$D_i = \frac{d_i}{\max_{i=1}^{20} d_i} \quad (3)$$

$$d_i = \sqrt{(x_i - x_p)^2 + (y_i - y_p)^2} \quad (4)$$

where $x_p$ and $y_p$ are the distances of the primary users and $x_i$ and $y_i$ are the distances of the secondary users from the primary user.

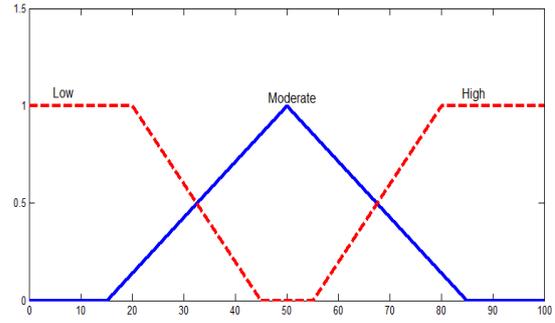

Fig. 2a. Membership Function(MF) used to represent the antecedent1

We apply different available spectrum inorder to find out the spectrum efficiency which is the main purpose of the opportunistic spectrum access strategy. Hence, we calculate the spectrum efficiency as the ratio of average busy spectrum over total available spectrum owned by secondary users.

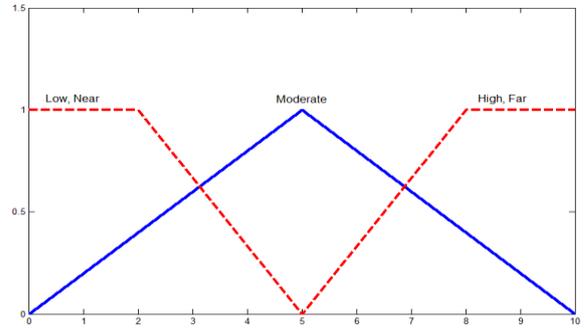

Fig. 2b. Membership Function(MF) used to represent the other antecedents

Linguistic variables are used to represent the spectrum utilization efficiency, distance and degree of mobility are divided into three levels: low, moderate, and high. while we use 3 levels, i.e., near, moderate, and far to represent the distance.

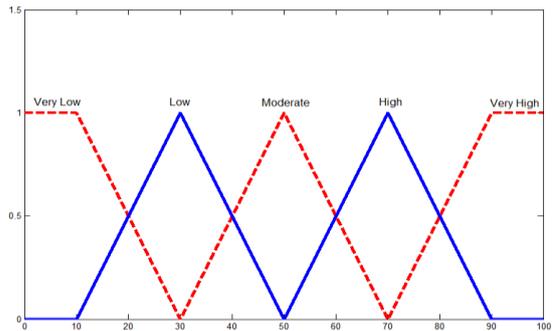

Fig. 2c. Membership Function(MF) used to represent the consequence

The consequence, i.e., the possibility that the secondary user is chosen to access the spectrum is divided into five levels which are very low, low, medium, high and very high. We use trapezoidal membership functions (MFs) to represent near, low, far, high, very low and very high, and triangle MFs to represent moderate, low, medium and high. MFs are shown in Fig. 2a, 2b, 2c. Since we have 3 antecedents and fuzzy subsets, we need setup $3^3 = 27$ rules for this FLS.

## V. KNOWLEDGE PROCESSING USING OPPORTUNISTIC SPECTRUM ACCESS

Table 1 summaries the various values based on the three antecedents and its consequence.

Table. 1
Opportunistic spectrum access using fuzzy rules

| Rule # | Antd.1 | Antd.2 | Antd.3 | Consq. |
|---|---|---|---|---|
| 1 | Low | Low | Near | Very Low |
| 2 | Low | Low | Moderate | Low |
| 3 | Low | Low | Far | Low |
| 4 | Low | Moderate | Near | Very Low |
| 5 | Low | Moderate | Moderate | Low |
| 6 | Low | Moderate | Far | Medium |
| 7 | Low | High | Near | Very Low |
| 8 | Low | High | Moderate | Low |
| 9 | Low | High | Far | Medium |
| 10 | Moderate | Low | Near | Very Low |
| 11 | Moderate | Low | Moderate | Medium |
| 12 | Moderate | Low | Far | High |
| 13 | Moderate | Moderate | Near | Very Low |
| 14 | Moderate | Moderate | Moderate | Medium |
| 15 | Moderate | Moderate | Far | High |
| 16 | Moderate | High | Near | Very Low |
| 17 | Moderate | High | Moderate | Low |
| 18 | Moderate | High | Far | High |
| 19 | High | Low | Near | Low |
| 20 | High | Low | Moderate | High |
| 21 | High | Low | Far | Very High |
| 22 | High | Moderate | Near | Low |
| 23 | High | Moderate | Moderate | High |
| 24 | High | Moderate | Far | Very High |
| 25 | High | High | Near | Very Low |
| 26 | High | High | Moderate | High |
| 27 | High | High | Far | High |

Since we chose a single consequent for each rule to form a rule base, we averaged the centroids of all the responses for each rule and used this average in place of the rule consequent centroid. Doing this leads to rules that have the following form:

$R'$: If Degree of mobility ($x_1$) is $F_l^1$; and its distance between primary user and the secondary users ($x_2$) is $F_l^2$; and the spectrum utilization efficiency of the secondary user ($x_3$) is $F_l^3$, then the Possibility (y) choosing the available spectrum is $c_{avg}^l$, where l = 1,2,..27 and $c_{avg}^l$ is defined as follows:

$$c_{avg}^l = \frac{\sum_{i=1}^{5} w_i^l c^i}{\sum_{i=1}^{5} w_i^l} \quad (5)$$

which $w_i^l$ is the number of choosing linguistic label i for the consequence of rule l and $c^i$ is the centroid of the i$^{th}$ consequence set (i: 1; 2; ...; 5; l: 1; 2; ...; 27). Table 2 provides $c^i$ for each rule. For every input ($x_1, x_2, x_3$), the output y($x_1, x_2, x_3$) of the designed FLS is computed as

$$y(x_1,x_2,x_3) = \frac{\sum_{i=1}^{27} \mu_{F_1^l}(x_1)\mu_{F_2^l}(x_2)\mu_{F_3^l}(x_3) c_{avg}^l}{\sum_{i=1}^{27} \mu_{F_1^l}(x_1)\mu_{F_2^l}(x_2)\mu_{F_3^l}(x_3)} \quad (6)$$

Table. 2
$c_{avg}^l$ to each corresponding rule

| Rule ≠ | $c_{avg}^l$ |
|---|---|
| 1 | 28.59 |
| 2 | 25.90 |
| 3 | 24.23 |
| 4 | 22.43 |
| 5 | 22.98 |
| 6 | 24.68 |
| 7 | 16.95 |
| 8 | 19.70 |
| 9 | 22.06 |
| 10 | 43.08 |
| 11 | 40.20 |
| 12 | 38.98 |
| 13 | 40.89 |
| 14 | 38.47 |
| 15 | 39.16 |
| 16 | 36.50 |
| 17 | 34.15 |
| 18 | 40.26 |
| 19 | 58.62 |
| 20 | 55.12 |
| 21 | 54.75 |
| 22 | 56.99 |
| 23 | 53.81 |
| 24 | 53.92 |
| 25 | 54.05 |
| 26 | 53.72 |
| 27 | 52.12 |

which gives the possibility that a secondary user is selected to access the available spectrum.

The weighted average value for each rule is given in the Table 2. Table 3 gives the three Descriptors and Possibility for four Secondary users. As listed in Table 3, at a particular time, values of three descriptors and possibility for four secondary user i.e., the secondary user chosen to the access the available spectrum is (SU4), the secondary user with the highest spectrum utilization is (SU2), the secondary user having the furthest distance to the primary user is (SU3), and the secondary user with the lowest mobility is (SU4).

Table 3.
Three Descriptors and Possibility for Four Secondary users

| Parameters | SU1 | SU2 | SU3 | SU4 |
|---|---|---|---|---|
| Distance from Primary user to secondary users | 8.01 | 2.16 | 12.80 | 6.02 |
| Degree of Mobility | 6.6667 | 9.2528 | 5.2229 | 1.2420 |
| Spectrum Utilization Efficiency | 61.0014 | 83.0274 | 70.8922 | 90.4191 |
| Possibility | 28.59 | 19.70 | 40.89 | 58.62 |

From table 3, we see that SU1 has 61%, SU4 has 90% of Spectrum utilization efficiency, SU3 only achieves 70.89% and SU4 achieves 90% of spectrum Utilization. Similarly, if we consider the Distance of primary user to the secondary users, SU1 has 8.01, SU2 has 2.16, SU3 has the furthest distance from the primary user 12.80 and SU4 has 6.02. Even though, if SU3 has furthest distance from the primary user but it has lowest spectrum utilization (70.89%) and lowest degree of mobility. Thus, the secondary user will select the spectrum for accessing based on the highest possibility rather than the highest spectrum utilization and the furthest distance from the primary user.

## VI. SIMULATION RESULTS

In this section, we present simulation results on the performance of our proposed work based on Fuzzy logic System. In the proposed work, we are choosing the available channel with the high possibility and high spectrum utilization efficiency.

To validate our approach, we randomly generated 20 secondary users over an area of 100 X 100 meters. The primary user was placed randomly in this area. Three descriptors were randomly generated for each secondary user. More specifically, the spectrum utilization efficiency of each secondary user was a random value in the interval [0,100] and its mobility degree in [0,10]. Distances to the primary users were normalized to [0,10].

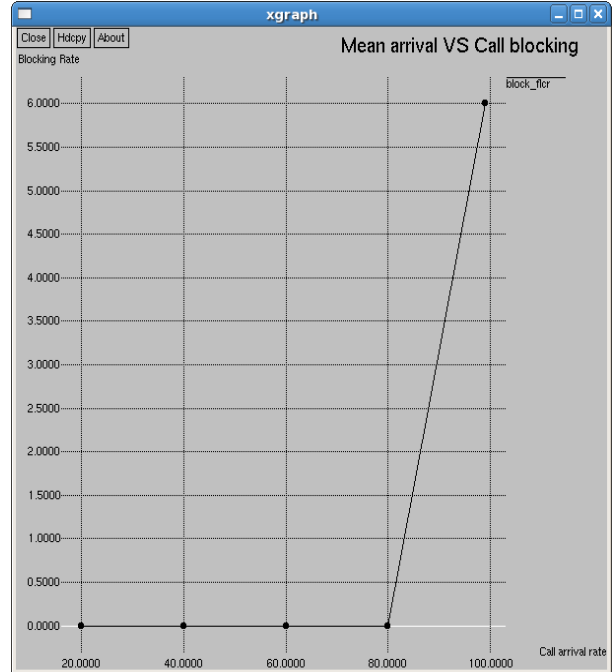

Fig. 3: Mean Arrival VS Call blocking rate

Fig. 3 shows the call blocking of the service provider using the Fuzzy logic system. As the call arrival rate increases the blocking rate gets decreased. Traffic rate increases along with the call blocking rate.

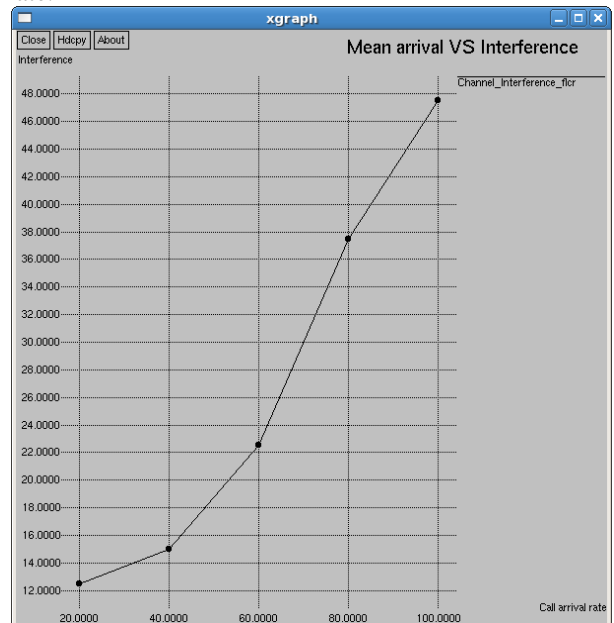

Fig. 4: Mean Arrival VS Interference using FLS

When interference increases spectrum utilization will decrease. The Fig. 4 shows that there is a decrease in the interference which provides an increase spectrum utilization using FLS. This leads to an efficient spectrum utilization.

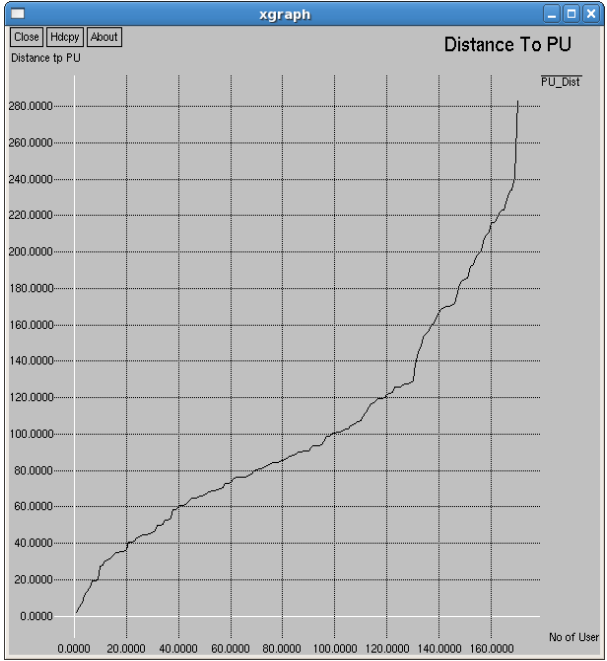

Fig. 5: Distance of Secondary Users to the Primary users using FLS

As illustrated in the Fig. 5, the distance from primary user to the secondary users is shown. The distance between primary user and the secondary users helps us for calculating the possibility for accessing the spectrum opportunistically.

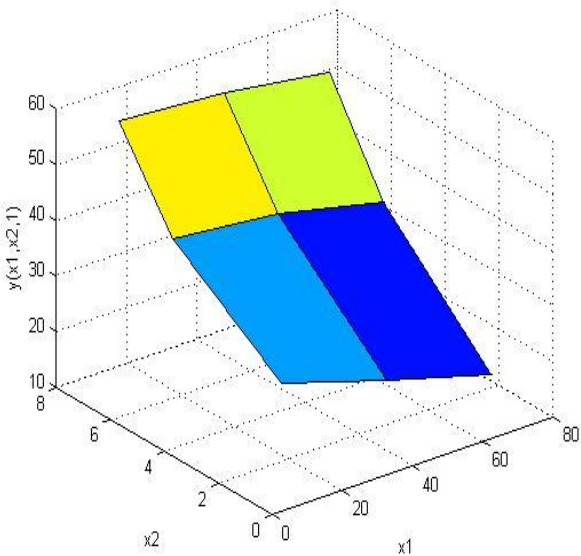

Fig. 6: The opportunistic spectrum access decision surface for the cognitive user with a fixed distance to the primary user: when the distance to the PU $x_3 = 1$.

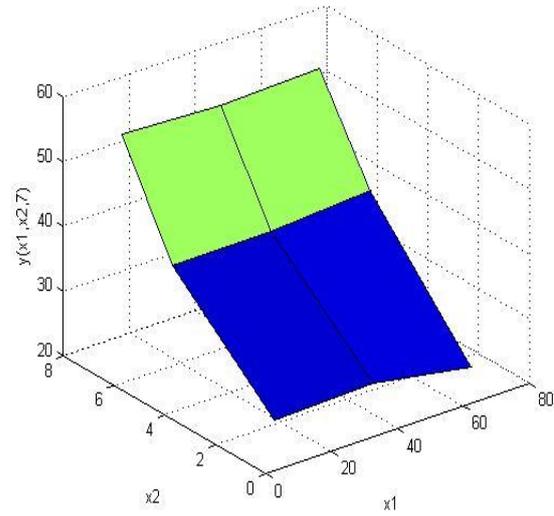

Fig. 7: The opportunistic spectrum access decision surface for the cognitive user with a fixed distance to the primary user: when the distance to the PU $x_3 = 7$.

Since it is impossible to plot visually, we fix one of three variables of y($x_1, x_2, x_3$). More specifically, we fixed the distance to the primary user $x_3$. Two cases, i.e., $x_3 = 1$ and $x_3 = 7$, were considered. Figure 6 & 7 represents the opportunistic spectrum access decision surface for the cognitive user for these two cases.

From Fig. 6&7, we see clearly that, at the same spectrum utilization efficiency and mobility degree, secondary users further from the primary user have higher chance to access the spectrum.

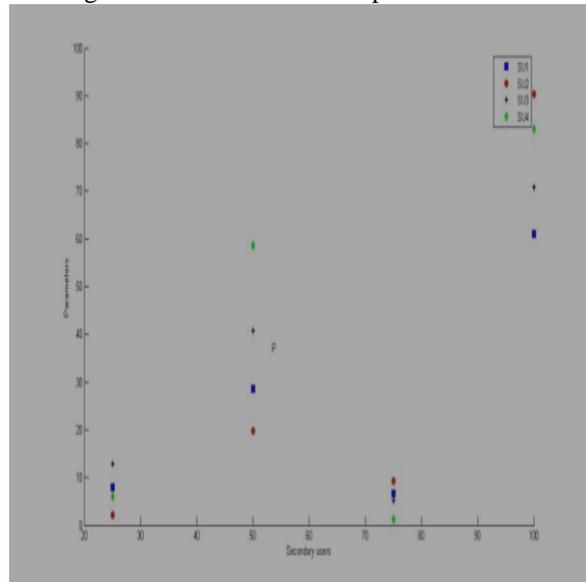

Fig. 8: An opportunistic spectrum access scenario in a specific space and a particular time.

The Fig. 8 illustrates that opportunistic spectrum access in a specific time and a particular time. Here, SU1, SU2, SU3 and SU4 are denoted by using the symbols ■ , ● , ✚ , ●. Here the Primary user is denoted by the symbol P.

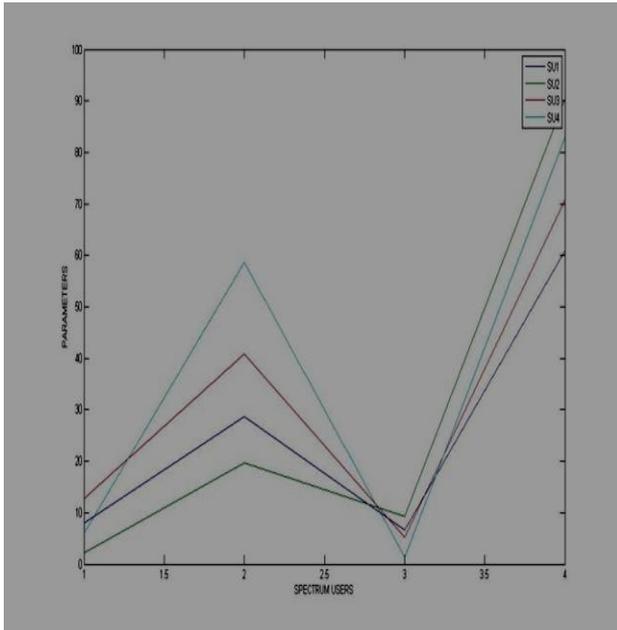

Fig. 9: Secondary Users with their corresponding parameters

In Fig. 9 shows that the Spectrum user (SU4) having the highest possibility of accessing the spectrum than the other users. Even though the other secondary users have the furthest distance from primary user to the secondary users, highest spectrum utilization and highest mobility but we prefer SU4 to access the spectrum because it has the highest possibility i.e., 58.62.

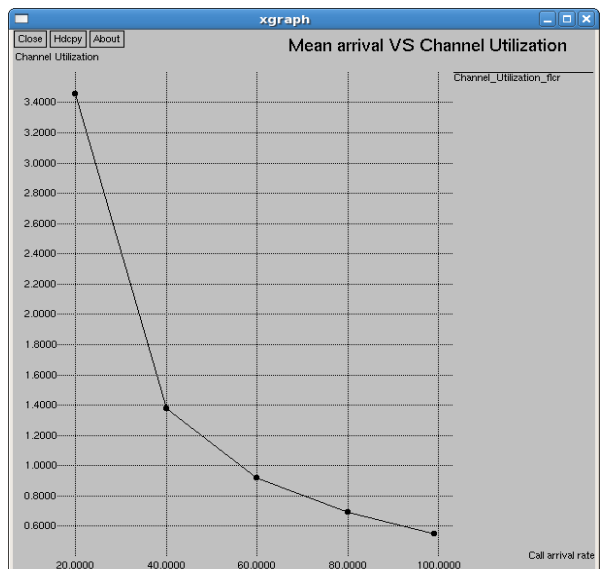

Fig. 10: Mean Arrival VS Channel Utilization using FLS

The Spectrum Efficiency (Channel Utilization) is defined as the ratio of average busy channels over total channels owned by service providers. The Fig. 10 shows that there is an increase in channel utilization with decrease in the call arrival rate.

Our designed FLS is used to control the spectrum assignment and access procedures in order to prevent multiple users from colliding in overlapping spectrum portions. Therefore, the secondary user with the highest possibility is guaranteed to access the spectrum.

## VII. CONCLUSION

This papers shows an efficient way to utilize the unused spectrum of the licensed users and it allows also other next generation mobile network users to benefit from available radio spectrum, leading to the improvement of overall spectrum utilization and maximizing overall PU and SU networks capacity.

We proposed an approach using a Fuzzy Logic System to detect the maximum possibility of spectrum access for secondary users via cognitive radio. The secondary users are selected on the basis of spectrum utilization, degree of mobility and distance from secondary users to the primary user. By using the above antecedents, we have calculated the highest possibility of accessing the spectrum band for secondary users and we have minimized the call blocking and interference. So that we can have better and efficient spectrum utilization.